\documentclass[12pt]{article} 
\usepackage{graphicx}
\usepackage{epsfig}
\usepackage{amsbsy}

\begin{document}
\baselineskip 10mm

\centerline{\large \bf Decay and fusion as two different
mechanisms} \centerline{\large \bf of stability loss for the
(C$_{20}$)$_2$ cluster dimer}

\vskip 6mm

\centerline{L. A. Openov$^{*}$ and A. I. Podlivaev}

\vskip 4mm

\centerline{\it Moscow Engineering Physics Institute (State
University), 115409 Moscow, Russia}

\vskip 2mm

$^{*}$ E-mail: LAOpenov@mephi.ru

\vskip 8mm

\centerline{\bf ABSTRACT}

The thermal stability of the (C$_{20}$)$_2$ cluster dimer
consisting of two C$_{20}$ fullerenes is examined using a
tight-binding approach. Molecular dynamics simulations of the
(C$_{20}$)$_2$ dimer at temperatures $T=$ 2000 - 3500 K show that
the finite lifetime $\tau$ of this metastable system is determined
by two fundamentally different processes, the decay of one of the
C$_{20}$ fullerenes and the fusion of two C$_{20}$ fullerenes into
the C$_{40}$ cluster. The activation energies for these processes
$E_a\approx$ 3.4 and 2.7 eV, respectively, as well as their
frequency factors, have been determined by analyzing the
dependence of $\tau$ on $T$.

\newpage

In 2000, after the experimental detection of the smallest
fullerene C$_{20}$ [1], Fig. 1a, on the surface of which C-C bonds
form only pentagons, whereas hexagons are absent (in contrast to
C$_N$ fullerenes with $N > 20$), a problem of synthesizing the
condensed phase (fullerite) based on this fullerene by analogue
with the known fullerite based on C$_{60}$ fullerenes [2, 3]
arised. Interest in this problem is stimulated, because the
C$_{20}$ fullerite may be a high-temperature superconductor [4-6].
Indeed, preliminary calculations point to a possibility of
formation of three-dimensional crystals consisting of C$_{20}$
fullerenes [4-6]. There is yet disagreement on the crystal
structure of such a hypothetical cluster matter [5,6], but all
authors agree that C$_{20}$ clusters in it should be connected
with each other by strong covalent bonds. This is a fundamental
difference of the C$_{20}$ fullerite from the C$_{60}$ fullerite
in which clusters are bonded with each other due to a weak van der
Waals attraction.

The strong intercluster interaction can significantly distort the shape of the C$_{20}$ clusters in the fullerite with
respect to the spherical shape of the C$_{20}$ fullerene and, as a consequence, result in the transition to another
crystalline (or even amorphous) modification. It seems to be reasonable to begin the analysis of the stability of the
C$_{20}$ fullerite with the analysis of the stability of the simplest system consisting of two interacting C$_{20}$
fullerenes, the (C$_{20}$)$_2$ cluster dimer. Experimental indications of the existence of charged (C$_{20}$)$_2^+$
dimers [as well as complexes (C$_{20}$)$_k^+$with $k=$ 3 - 13] were obtained in [7]. The dimerization of neutral
fullerenes C$_{20}$ was theoretically studied in [4] and, most comprehensively, in [8], where eight different
(C$_{20}$)$_2$ isomers were found, their structural and energy characteristics were calculated, and it was shown that
the so-called {\it open}-[2+2] isomer has the lowest energy, see Fig. 1b (numbers in the square brackets are the numbers
of the atoms of each fullerene that are involved in the formation of intercluster bonds).

The aim of this work is to theoretically analyze the stability of
the (C$_{20}$)$_2$ dimer with respect to its transition to another
atomic configuration. It is known (see, e.g., [9]) that the
specific (per atom) binding energy of carbon clusters, which is
defined as the difference between the energy of an isolated carbon
atom and the specific potential energy of the
$N$-atomic cluster, $E_b(N)/N=E(1)-E_{pot}(N)/N$, increases with
$N$ (this is the property of the overwhelming majority of the
elements of the Periodic Table, except for metastable clusters of
nitrogen, helium, and some other elements [10, 11]). For this
reason, it is energetically favorable for two (or several) carbon
clusters to merge into one larger cluster, thereby increasing the
binding energy [10] (i.e., reducing the potential energy), by
analogy with fusion of light atomic nuclei. At the same time, our
previous analysis [12] of the thermal stability of the isolated
C$_{20}$ fullerene showed that, when heated, it passes to various
energetically unfavorable configurations with lower $E_b$ values
rather than to a more stable isomer with the maximum binding
energy ("bowl"). This is due to the ability of carbon structures
to form numerous "intermediate" metastable states separated by
high energy barriers from configurations with a lower potential
energy.

Since the C$_{20}$ fullerenes in the (C$_{20}$)$_2$ dimer are, on
the one hand, at small distances from each other and hence strongly
interact with each other and, on the other hand, both preserve
the overall spherical shape (see Fig. 1b), the (C$_{20}$)$_2$ dimer
can loose its "molecular" structure both through the fusion of two
C$_{20}$ fullerenes to the C$_{40}$ cluster and through the decay
of one of them [12]. A question arises of which of two factors,
thermodynamic (the minimum of the energy of the final product
after fusion) or kinetic (the thermally activated transition to
the metastable configuration with a high energy upon decay),
finally determines the stability of the (C$_{20}$)$_2$ dimer? To
answer this question, we study the dynamics of the (C$_{20}$)$_2$
dimer numerically at high (sufficient to overcome the
corresponding potential barriers $U\sim 5$ eV [12]) temperatures.

To calculate the energies of arbitrary atomic configurations, we
used the tight binding model [13], which is a reasonable
compromise between {\it ab initio} calculations and too simplified
empirical approaches with classical inter-atomic potentials. This
model well describes both small clusters and macroscopic carbon
structures. Previously, we used it to simulate the fullerenes
C$_{20}$, C$_{60}$, and other carbon systems [10, 12, 14-18]. For
the binding energy of an isolated C$_{20}$ fullerene, this method
gives $E_b=6.08$ eV/atom [12] (in agreement with a value of
$E_b=6.36$ eV/atom obtained in {\it ab initio} calculations [19).
For the binding energy of the {\it open}-[2+2] isomer of the
(C$_{20}$)$_2$ dimer, see Fig. 1b, this method gives a value of
$E_b=6.20$ eV/atom (which corresponds to the coagulation energy of
two fullerenes $\Delta
E=2E_{pot}$[C$_{20}$]-$E_{pot}$[(C$_{20}$)$_2$] = 4.9 eV, to be
compared with the value $\Delta E=6.3$ eV obtained by the density
functional method [8]).

To analyze the thermal (it can also be called kinetic) stability of the (C$_{20}$)$_2$ dimer, we study its dynamics
numerically for the initial temperature $T_{ini}=$ 2000 - 3500 K and fixed total energy $E_{pot}+E_{kin}=$ const, where
$E_{kin}$ is the kinetic energy in the center-of-mass syatem. In this formulation of the problem, the system temperature
$T$ is a measure of the energy of the relative motion of atoms (for more details, see [12, 17, 18]). The values of $T$
were calculated by the formula $\langle E_{kin}\rangle = \frac{1}{2}k_BT(3N-6)$, where $k_B$ is the Boltzmann
constant, $N=40$ is the number of atoms in the (C$_{20}$)$_2$ dimer, and the angular brackets stand for averaging
over $10^3 - 10^4$ steps of the molecular dynamics simulation (the time of one step is $t_0=2.72\cdot10^{–16}$ s).

We observed two fundamentally different types of the $T(t)$
dependence. They are shown in Fig. 2. In the first case (Fig. 2a),
the dimer temperature at $t\approx 4.4$ ns drops stepwise by
$\Delta T\approx$ 200 K. The decrease in temperature points to a
decrease in the kinetic energy and, hence, an increase
in the potential energy (because $E_{pot}+E_{kin}=$ const).
Therefore, the (C$_{20}$)$_2$ dimer passes to a less favorable
configuration. Our analysis showed that this transition is caused
by the decay of one of the C$_{20}$ fullerenes, whereas the other
fullerene preserved the cage shape, see Fig. 3a. The binding
energy of the metastable state formed upon the decay is equal to
(after the relaxation to the local minimum of the potential
energy) $E_b=6.14$ eV/atom, which is lower than the value of $E_b$
for the initial (C$_{20}$)$_2$ dimer by $\Delta E_b=0.06$ eV/atom.
This $\Delta E_b$ value is consistent with the value $\Delta
T\approx$ 200 K, taking into account that, in thermal oscillations
near the local minimum of the potential energy $E_{pot}^{min}$ ,
the average kinetic energy of the system is approximately (up to
anharmonic corrections) equal to the average increase in the
potential energy $\Delta E_{pot}=E_{pot}-E_{pot}^{min}$, so that,
since the total energy is constant, i.e.,
$E_{pot}+E_{kin}=E_{pot}^{min}+\Delta E_{pot}+E_{kin}\approx
E_{pot}^{min}+2E_{kin}\approx E_{pot}^{min}+3Nk_BT$, in the
$1\rightarrow 2$ transition from the vicinity of one local minimum
to the vicinity of the other local minimum, the relation
$E_{pot}^{min,1}+3Nk_BT_1\approx E_{pot}^{min,2}+3Nk_BT_2$ is
valid, from which it follows that $\Delta T=T_1-T_2\approx \Delta
E_b/3k_B$ for $E_{pot}^{min,2}>E_{pot}^{min,1}$ taking into
account the definition of $E_b$.

On contrary, for the second case (Fig. 2b), the temperature at
$t\approx 4.4$ ns increases stepwise by $\Delta T\approx 200$ K.
This indicates an increase in the kinetic energy, i.e., a decrease
in the potential energy. Therefore, the system passes to the
energetically favorable configuration. However, since the initial
{\it open}-[2+2] isomer has the minimum energy among all
(C$_{20}$)$_2$ dimers, this transition indicates that the C$_{20}$
fullerenes loss their individuality and merge into the C$_{40}$
cluster. We analized the atomic structure after the temperature
jump and showed that this is indeed the case. The C$_{40}$ cluster
formed upon fusion is shown in Fig. 3b. It can be considered as
one of nonclassical, strongly imperfect isomers of the C$_{40}$
fullerene: the C-C bonds in it form not only pentagons and
hexagons, but also two 9-member rings. The binding energy of this
cluster after relaxation, $E_b=6.25$ eV/atom, is higher than the
value for the (C$_{20}$)$_2$ dimer by $\Delta E_b=0.05$ eV/atom,
which is consistent with a value of $\Delta T\approx 200$ K upon
fusion.

We also observed the fusion of the C$_{20}$ fullerenes into
isomers with higher binding energies. Fig. 4 shows the $T(t)$
dependence for the case, where an imperfect isomer of the C$_{40}$
fullerene with one heptagon and $E_b=6.49$ eV/atom is formed upon
fusion at $t\approx 0.34$ ns. At $t\approx 0.46$ ns, the
annealing of the defect takes place and the C$_{40}$ fullerene
with $E_b=6.53$ eV/atom is formed, on the surface of which the C-C
bonds form 12 pentagons and 10 hexagons, see Fig. 5.
In this case, the temperature increases
sequentially by $\Delta T\approx 1100$ and 150 K, which is
consistent with changes $\Delta E_b$, see above. Transitions of
the (C$_{20}$)$_2$ dimer into other isomers of the C$_{40}$
fullerene were also observed. The general statistics is as follows
($N_k$ is the number of $k$-gons with $k > 6$ in the C$_{40}$
isomer). Among 57 sets of the initial displacements and velocities
of atoms that correspond to different initial
temperatures $T_{ini}$, the fusion into the C$_{40}$ cluster with
$N_9=2$ occurs most frequently (16 times), see Fig. 3b. In
addition, the clusters with $N_{10}=2$ and $N_8=1$ (six times);
$N_9=N_8=N_7=1$ (once); $N_8=2$ (twice); $N_8=N_7=1$ (once);
$N_7=1$ (once); and $N_7=1$ (once) are formed. In other 29 cases,
the decay of one of the C$_{20}$ fullerenes takes place (Fig. 3a). The
further evolution of the system after it looses its molecular
structure is as follows. The decay of one of the fullerenes in the
(C$_{20}$)$_2$ dimer is followed by the decay of the second
fullerene, while fusion of two fullerenes into the C$_{40}$
cluster is followed by the decay of this cluster, so that the
final product is always a quasi-two-dimensional or
quasi-one-dimensional configuration as in the decay of the
isolated C$_{20}$ fullerene [12]. We never observed the return of
the system to the initial {\it open}-[2+2] configuration. The
fission of the (C$_{20}$)$_2$ dimer into two isolated C$_{20}$
fullerenes has not been observed as well.

There is correlation between the value of $T_{ini}$ and the
mechanism of the loss of the stability of the (C$_{20}$)$_2$
dimer. The loss of the molecular structure of the dimer occurs
primarily due to fusion of two C$_{20}$ fullerenes into the
C$_{40}$ cluster at $T_{ini} < 2500$ K, due to the decay of one of
the fullerenes at $T_{ini} > 3000$ K, and due to both those
processes at 2500 K $< T_{ini} < 3000$ K, see Fig.6. To determine
the activation energies $E_a$ for the decay and fusion, we
analyzed the temperature dependence of the lifetime $\tau$ of the
(C$_{20}$)$_2$ dimer, using the Arrhenius formula with the
finite-heat-bath correction [18, 20, 21]
\begin{equation}
\tau^{-1}(T)=A\cdot\mathrm{exp}\left[-\frac{E_{a}}{k_{B}T\left(1-E_{a}/2CT\right)}\right].
\end{equation}
Here, $A$ is the frequency factor, $C$ is the microcanonical
specific heat, which we set to $C=(3N-6)k_B$, where $N=40$ (as
will be shown below, $E_a/2CT\approx 0.05 << 1$ for $T\approx
3000$ K). Figure 6 shows the calculated logarithm of $\tau$ as a
function of ($T_{ini}^{*}$)$^{-1}$, where
$T_{ini}^{*}=T_{ini}(1-E_a/2CT_{ini})$, see Eq. (1). It is seen
that this dependence in a quite wide range $T_{ini} = 2000 - 3000$
K is given, in the first approximation, by the straight line for
both the decay and fusion processes; i.e., $A(T)\approx$ const
(since the spread of the points is sufficiently large,
particularly for fusion processes, it is impossible to reliably
determine the temperature dependence of $A$). Using the slopes of
these straight lines and the points of their intersection with the
ordinate axis, we determined the activation energies $E_a=3.4\pm
0.2$ eV and $E_a=2.7\pm 0.3$ eV, as well as the frequency factors
$A=(1.8\pm 0.3)\cdot 10^{17}$ s$^{-1}$ and $A=(5.6\pm 1.1)\cdot
10^{15}$ s$^{-1}$ for the decay and coalescence processes,
respectively. [Since $E_a$ enters both into the nominator of the
exponential in Eq. (1) and into the "renormalized" initial
temperature, it is determined self-consistently by successive
iterations]. The values found for $E_a$ are one fourth to one
third of the fragmentation activation energy of the C$_{60}$
fullerene (frequency factors are smaller by two to four orders of
magnitude) [18].

In conclusion, we note that the value $E_a\approx 3.4$ eV for the decay of one C$_{20}$ fullerene in the (C$_{20}$)$_2$
dimer is approximately one half of the value $E_a\approx 7$ eV for the decay of the isolated C$_{20}$ fullerene [12].
This fact, along with the appearance of the second mechanism of the loss of the molecular structure (fusion of two C$_{20}$
fullerenes into the C$_{40}$ cluster) with a lower value of $E_a\approx 2.7$ eV, indicates that stability decreases
when the (C$_{20}$)$_2$ dimer is formed from two C$_{20}$ cages.
In three-dimensional cluster structures based on C$_{20}$ fullerenes, both the stabilization of the metastable state due to an
increase in the number of nearest neighbors of each C$_{20}$ fullerene and the appearance of new channels of
stability loss can be expected. We are going
to study this problem in the next work.



\newpage

\includegraphics[width=\hsize,height=19cm]{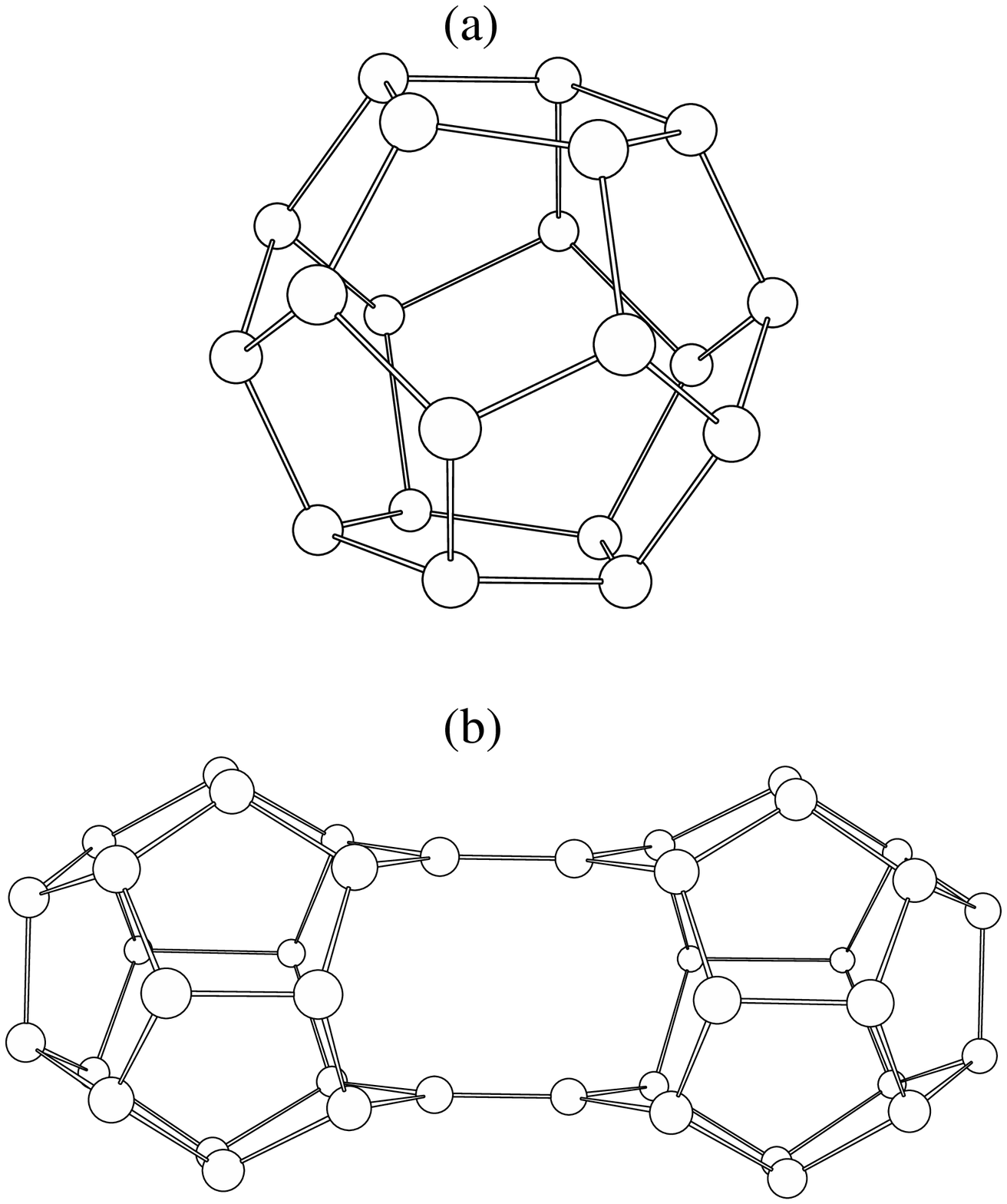}

\vskip 5mm

Fig. 1. (a) C$_{20}$ fullerene with the binding energy $E_b=6.08$ eV/atom and (b) (C$_{20}$)$_2$
cluster dimer with the binding energy $E_b=6.20$ eV/atom.

\newpage

\includegraphics[width=10cm,height=17cm]{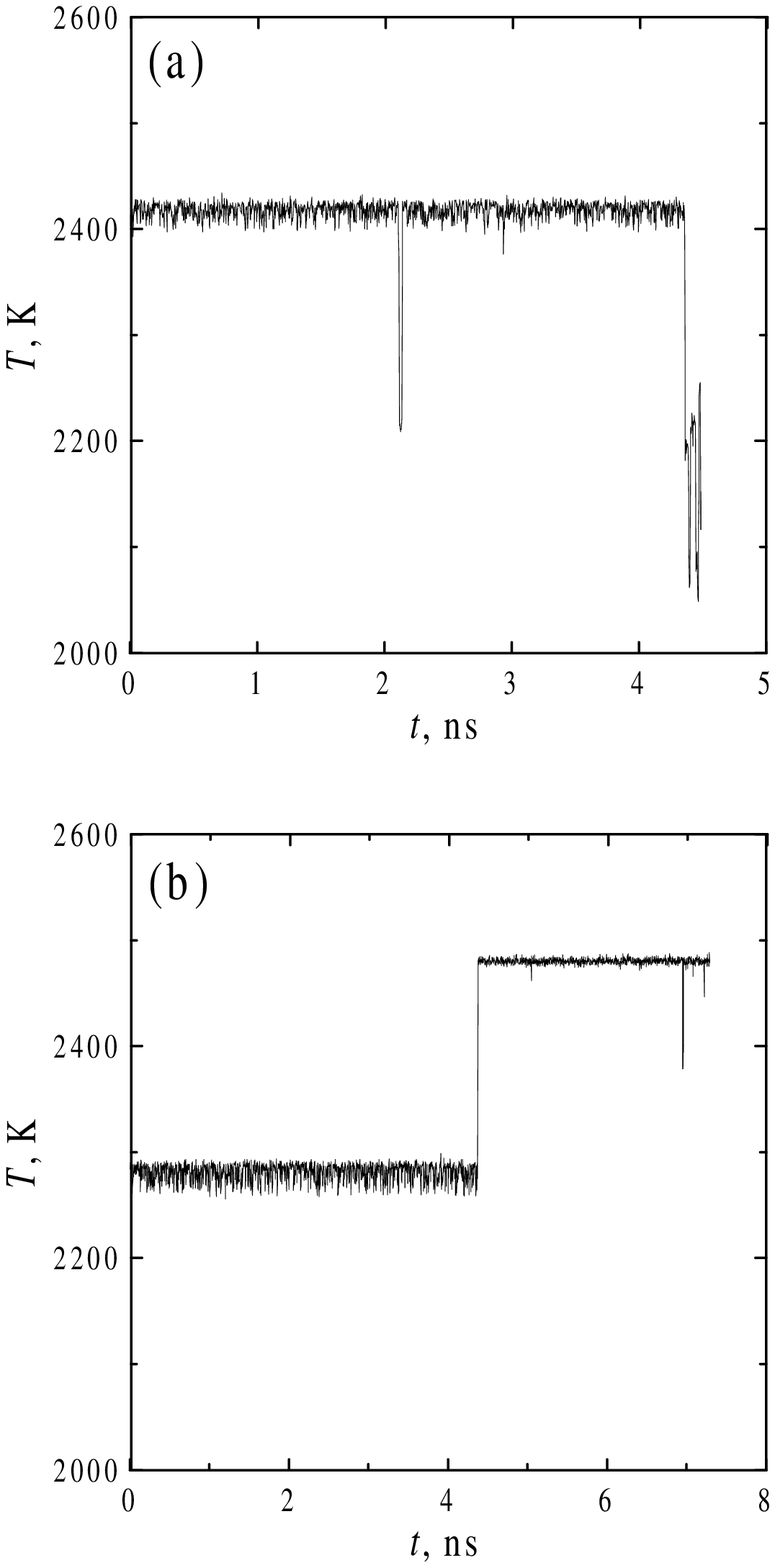}

\vskip 5mm

Fig. 2. Time dependences of the temperature $T$ of the (C$_{20}$)$_2$ cluster dimer obtained from the molecular
dynamics simulation by averaging the kinetic energy over each $10^4$ steps. The total energy of the system is constant.
The time of one step is equal to $t_0=2.72\cdot10^{-16}$ s. The initial temperature is $T_{ini}=2415\pm 10$ K (a) and
$2285\pm 10$ K (b).

\newpage

\includegraphics[width=13cm,height=18cm]{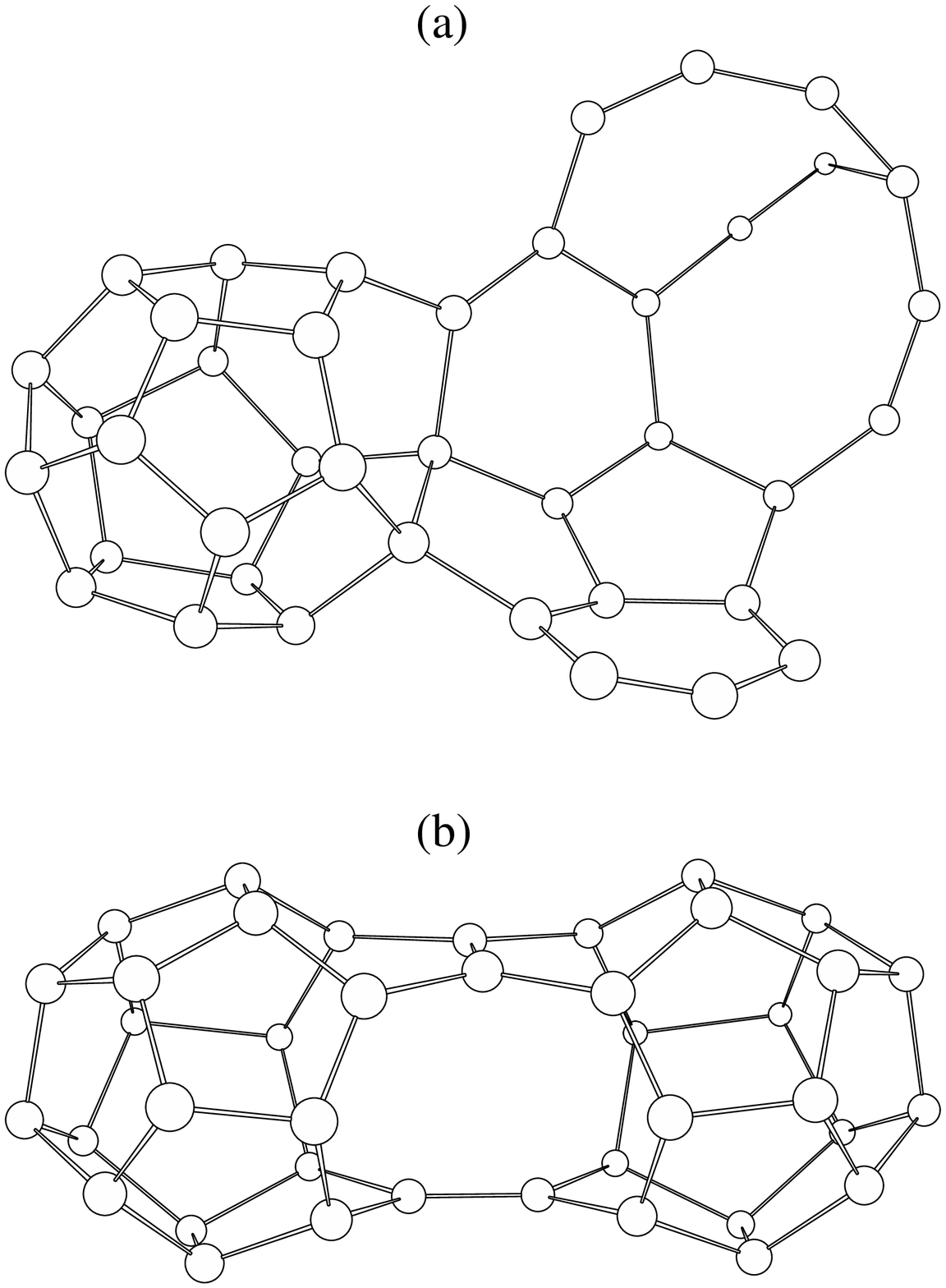}

\vskip 5mm

Fig. 3. (a) Atomic configuration after the decay of one of the C$_{20}$ fullerenes in the (C$_{20}$)$_2$ dimer, see Fig. 2a.
The binding energy after relaxation is $E_b=6.14$ eV/atom. (b) The C$_{40}$ cluster formed upon fusion of two C$_{20}$
fullerenes in the (C$_{20}$)$_2$ dimer, see Fig. 2b. The binding energy after relaxation is $E_b=6.25$ eV/atom.

\newpage

\includegraphics[width=15cm,height=14cm]{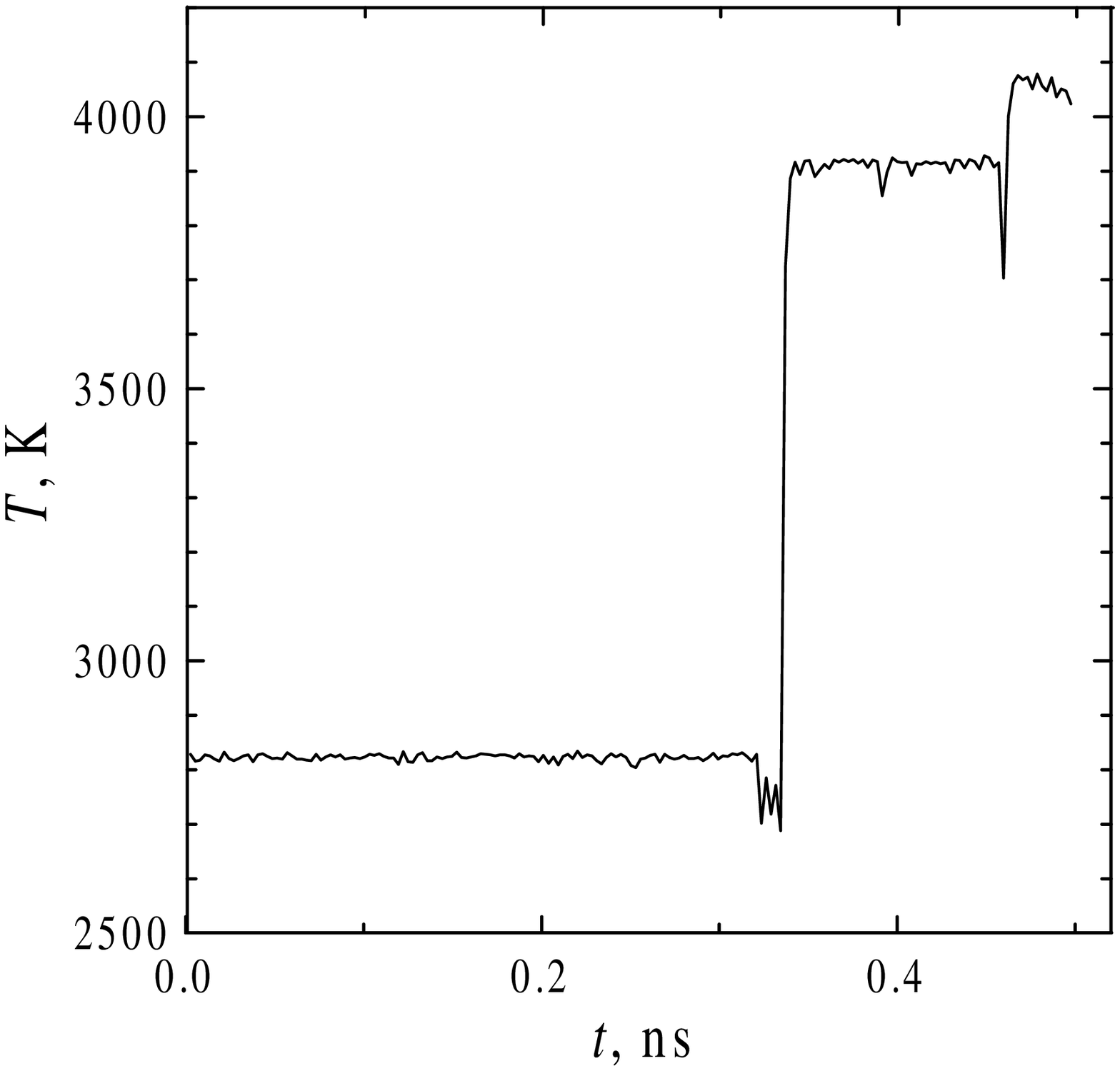}

\vskip 5mm

Fig. 4. Same as in Fig. 2, for $T_{ini}=2820\pm 10$ K.

\newpage

\includegraphics[width=13cm,height=18cm]{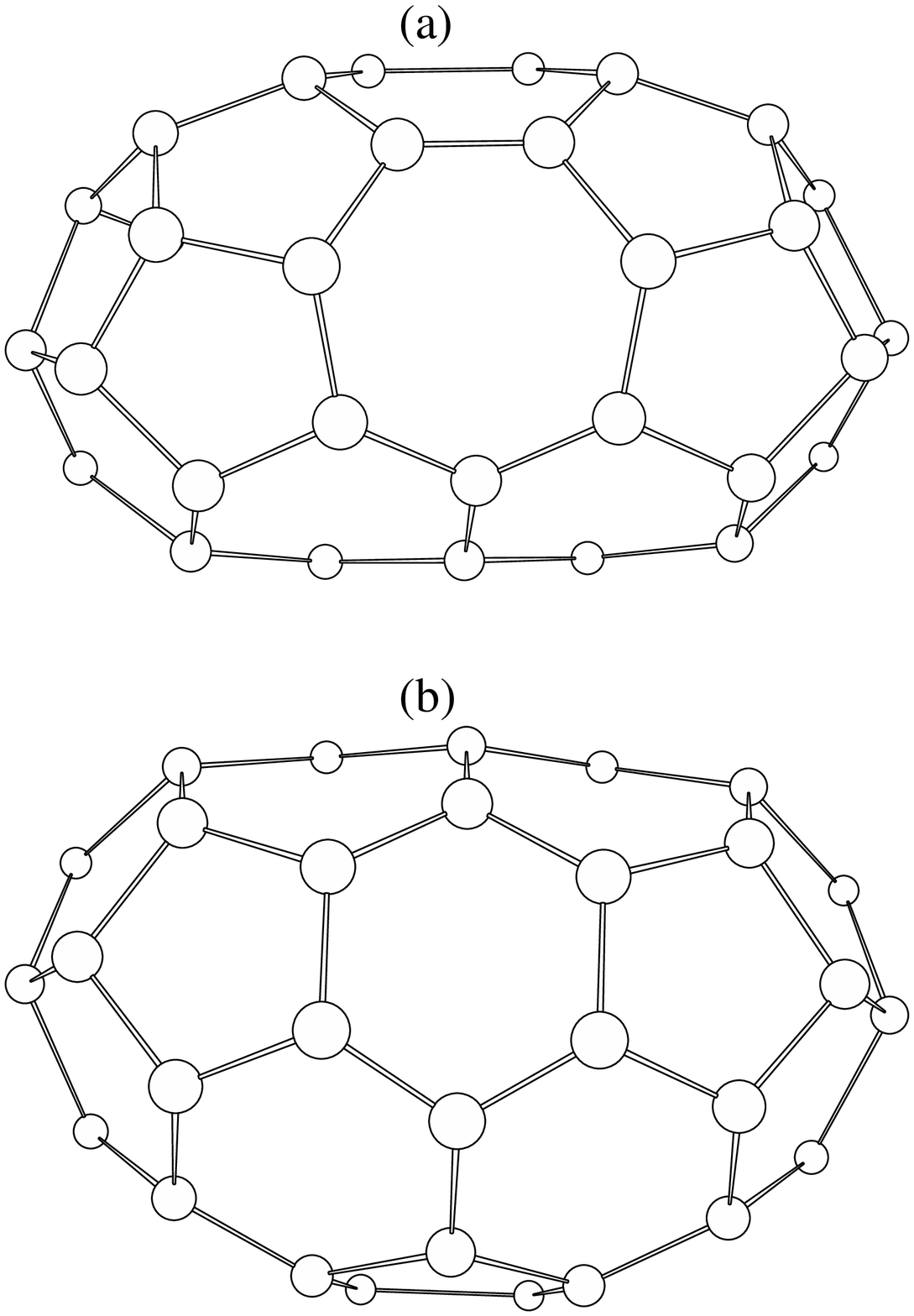}

\vskip 5mm

Fig. 5. (a) Defect isomer of the C$_{40}$ fullerene that is formed
upon fusion of two C$_{20}$ fullerenes in the (C$_{20}$)$_2$
dimer, see Fig. 4. The binding energy after relaxation is
$E_b=6.49$ eV/atom. (b) The C$_{40}$ fullerene formed after
annealing of the imperfect isomer. The binding energy after
relaxation is $E_b=6.53$ eV/atom. The distant atoms are not shown
for clarity.

\newpage

\includegraphics[width=16cm,height=15cm]{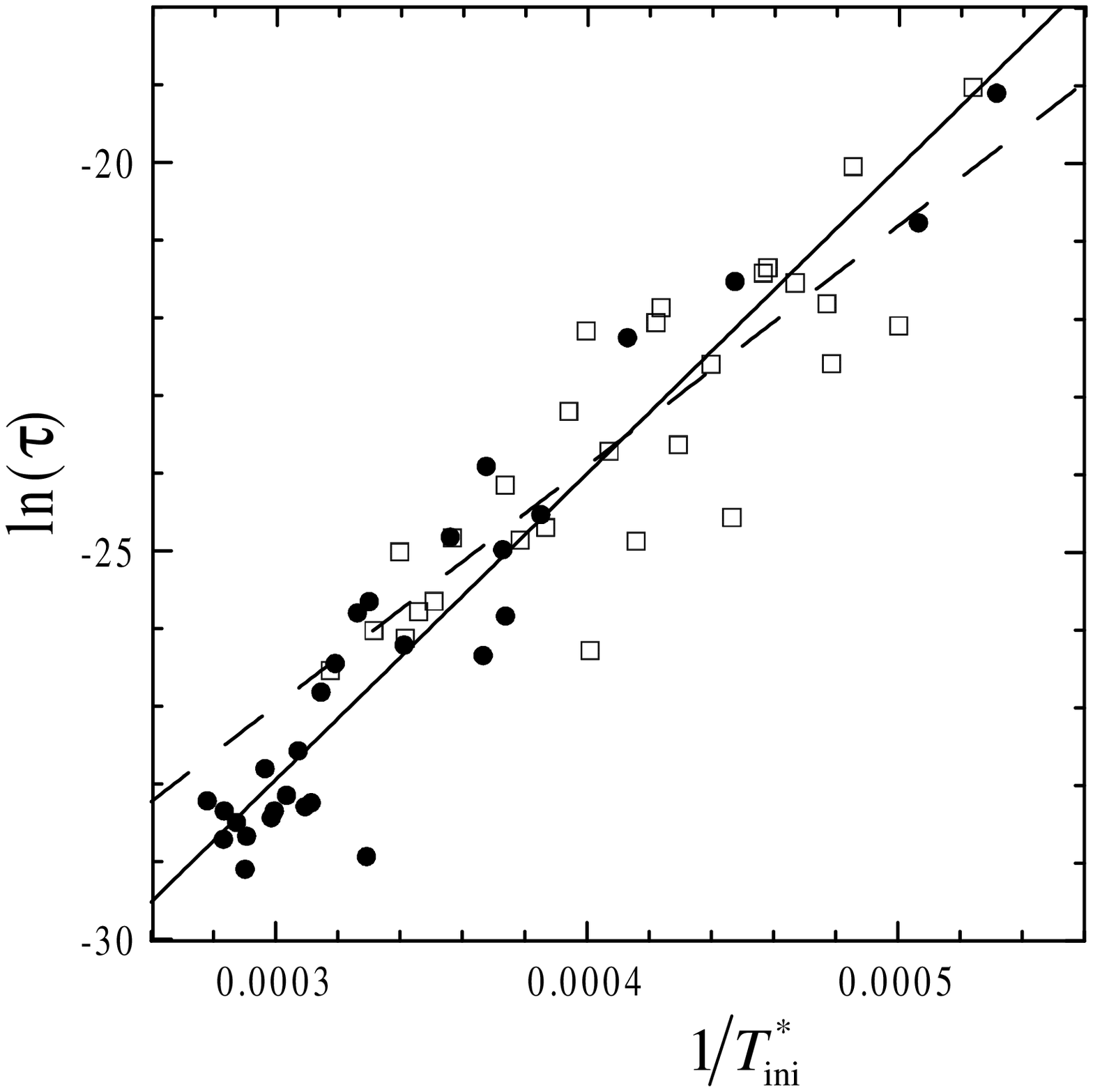}

\vskip 5mm

Fig. 6. Logarithm of the lifetime $\tau$ (in seconds) of the (C$_{20}$)$_2$ cluster dimer for the processes of the decay
of one of C$_{20}$ fullerenes (circles) or fusion of two C$_{20}$ fullerenes into the C$_{40}$ cluster (squares) versus the
inverse temperature (Kelvins), taking the finite-heat-bath correcrtion into account, see text. The solid and dashed lines
are the linear approximations for decay and fusion, respectively.

\end{document}